\documentclass[a4paper,11pt]{article}
\pdfoutput=1
\usepackage{jheppub}
\usepackage{amsmath}
\usepackage{amsthm}
\usepackage{amsfonts}
\usepackage{amssymb}
\usepackage{textcomp}
\usepackage{graphicx}% Include figure files
\usepackage{bm}% bold math
\usepackage{color}
\usepackage{verbatim}
\usepackage{subcaption}
\usepackage{graphicx}
\usepackage{epsfig}

\definecolor{orange}{rgb}{1,0.5,0}
\definecolor{col1}{RGB}{153, 52, 121}
\definecolor{dgreen}{rgb}{0,0.55,0}
\definecolor{pink}{rgb}{1,0.08,0.58}

\usepackage{amsfonts,amsmath,amssymb}

\newcommand{\p}{\partial}

\newcommand{\rar}{\rightarrow}

\begin{document}

 \title{
 \Huge Relaxation regimes of the holographic electrons at charge neutrality after a local quench of chemical potential.
 \color{black}}
 
 \author[a,b]{Alexander Krikun,\footnote{https://orcid.org/0000-0001-8789-8703}}

\affiliation[a]{Instituut-Lorentz for Theoretical Physics, $\Delta ITP$, Leiden University, Niels Bohrweg 2, Leiden 2333CA, The Netherlands} 

\affiliation[b]{Skolkovo Institute of Science and Technology, Nobel street 3, Moscow 121205, Russia
}

% e-mail addresses: one for each author, in the same order as the authors
\emailAdd{krikun@lorentz.leidenuniv.nl}

\abstract{
In this work we study the relaxation of the system of strongly correlated electrons, at charge neutrality, when the chemical potential undergoes a local change. This setup is a model for for the X-ray absorbtion edge study in the half-filled graphene. We use holographic duality to describe the system as a classical Schwarzschild black hole in curved 4-dimensional AdS spacetime. Assuming the amplitude of the quench is small, we neglect the backreaction on the geometry. We numerically study the two relaxation regimes: the adiabatic relaxation when the quench is slow and the relaxation governed by the quasinormal modes of the system, when the quench is fast. We confirm the expectation that the scale of separation between the slow and fast regimes is set by the characteristic frequency of the quasinormal modes.
}

\maketitle

\section{Introduction}
Quantum features of the macroscopic systems always attract a lot of interest and give rise to new unexpected phenomena which can in many cases lead to the development of useful quantum devices. The examples are superconductors, light-emitting diodes, quantum computers, etc. The development of the latter ones addresses among others the interesting problem of quantum non-equilibrium dynamics, which can demonstrate quite unusual features. 

The prominent example of the substantially quantum non-equilibrium behavior is the effect called orthogonality catastrophe (OC). Using the fundamental features of the weakly coupled quantum many-electron systems one can show that the response of such a system to the sudden introduction of the local impurity behave as a power law in time, which furthermore leads to the appearance of the power-law edge singularities in the X-ray absorption spectra \cite{mahan2013many}. This particular problem has recently gained more interest since the corresponding weakly interacting fermionic systems can now be realized with the cold atoms setups \cite{knap2012time,schmidt2017universal}. 

The standard treatment of the orthogonality catastrophe relies on the assumption that the fermionic degrees of freedom of the system (quasiparticles) can be described as a nearly free Fermi gas. The natural question arises therefore whether the similar phenomenon can be seen in the strongly correlated electronic systems, where the quasiparticle description in unavailable. A particular examples of such systems are srange metallic state of high temperature superconductors, or, as we focus on in this paper, the half filled state of graphene, which can be viewed as a system of strongly interacting Dirac fermions at charge neutrality point \cite{Crossno1058}. In order to address this question we invoke the AdS/CFT duality  \cite{Maldacena:1997re,Witten:1998qj,Gubser:1998bc} (holography) which fundamentally relates the strongly coupled quantum field theory (an exact opposite to the nearly-free Fermi gas) to the classical gravity in a curved spacetime with one extra dimension. 
In holographic framework the system of strongly correlated electrons at charge neutrality point in 2+1 dimensions at finite temperature is equivalent to the Schwarzschild black hole the 3+1 dimensional Anti-de-Sitter spacetime\footnote{In case of finite background chemical potential the corresponding black hole would be the charged Reissner-Nordstr\"om one}. The horizon radius of the black hole is controlled by the temperature of the strongly correlated theory.

In this work we start the investigation of the orthogonality catastrophe in strongly correlated quantum systems by first addressing the basic problem of the non-equilibrium evolution of the charge neutral model after a local quench in the chemical potential. At time $t=0$ the local deformation in the spatial profile of the chemical potential is created. This corresponds roughly to the X-ray absorption experiment, when the core hole is created and therefore the electrons in the conduction band feel the defect in the background potential. The present work, while being focused on a quite simple holographic setup, is relevant to the existing studies of X-ray anomaly in charge neutral graphene \cite{yang2007x}.

The non-equilibrium dynamics in the holographic framework is a well developed subject (see \cite{Liu:2018crr} for a recent review). Unlike quantum systems, the real time evolution of the dual gravitational model is relatively easy to address and some important results can be even obtained in exact analytic form. The global homogeneous quenches in the context of condensed matter applications have been studied in \cite{Bhaseen:2012gg} and the non-homogeneous stochastic deformations were addressed in \cite{Sonner:2014tca}. The evolution due to global quench of the chemical potential has been studied in \cite{Caceres:2012em,Caceres:2014pda}. The static configurations with a point-like deformation of the chemical potential have been constructed in \cite{Horowitz:2014gva}. The holographic models for the pump-probe experiments, closely related to the OC, have been explored in \cite{Withers:2016lft,Bagrov:2017tqn,Bagrov:2018wzu}. Clearly the holographic technology provides all the tools to address our problem. The extra technical novelty in our work in particular is the usage of the spherical coordinates on the boundary, which proves to be especially convenient in case of point-like excitation.  

We will focus on the simplest case when the perturbation of the chemical potential is small to the extent that the backreaction on the metric can be neglected and the problem reduces to the study of non-equilibrium dynamics of the U(1) gauge field in the background of AdS-Schwarzschild black hole. The convenient advantage of the holographic framework is that one can study the time evolution problems at finite temperature by directly solving the classical hyperbolic equations of motion. 

As mentioned above the characteristic feature of the OC is the power law approach to equilibrium state at late times after the quench. We study the different cases, changing the timescale at which the defect in the chemical potential develops. We observe that the adiabatic evolution, when the system follows the slow development of the defect, gets substituted by the exponential relaxation, in case of the fast quench, when the quasinormal modes (QNMs) of the system get excited. The transition between the two regimes happens roughly at the scale of the lowest QNM frequency, which is controlled by the temperature. In the particular example under consideration we do not observe the power law relaxation, which is characteristic to the OC. Quite interestingly, this observation is in line of the recent results of \cite{deng2015exponential}, where the exponential OC has been reported for quantum many-body localized systems. In the case presented here we attribute the exponential behavior to the effective linear structure of the equations of motion, which don't include gravity. However it is also especially interesting to figure out how this feature gets affected in the holographic systems at finite chemical potential, where the backreaction can not be omitted and the full nonlinear dynamics can be triggered. 

The paper is organized as follows. In Sec.\,\ref{sec:setup} we outline the holographic setup, then we discuss the non-equilibrium evolution after a local quench in Sec.\,\ref{sec:point_quench}. In Sec.\,\ref{sec:QNMs} we compute the spectrum of cylindrical quasinormal modes, which govern the late time evolution of the system in case of the fast quench. Finally, we conclude and discuss the future directions in Sec.\,\ref{sec:concl}. The two Appendices are devoted to the details of the numerical method of lines and the calculation of the quasinormal mode spectrum with numerical shooting method. 

\section{\label{sec:setup}Holographic model} 

In this work we focus on the simplest setup with the strongly interacting electrons in 2+1 dimensions at vanishing chemical potential. In the holographic framework this system is described via the Schwarzschild black hole in 3+1 dimensional curved AdS spacetime. In what follows we will adopt the polar coordinates on the boundary ($\rho$,$\phi$) therefore the metric reads 
\begin{gather}
\label{equ:RN}
ds^2 = \frac{L^2}{z^2} \left( - f(z) dt^2 + \frac{dz^2}{f(z)} + d\rho^2 + \rho^2 d\phi^2\right), \\
f(z) = \left(1-\frac{z^3}{z_h^3} \right), \qquad A_\mu(z) = 0.
\end{gather}
where $A_\mu$ is a gauge field dual to the $U(1)$ conserved charge of the fermion particle number, the $AdS$ boundary is located at $z=0$ and the black hole horizon is located at $z=z_h$. $L$~is the $AdS$ curvature radius which can be set to unity by means of conformal symmetry. The above metric is a solution to the classical equations of motion following from the Einstein-Maxwell action
\begin{equation}
\label{equ:s_EM}
S = \int \sqrt{g} \left[R - 2 \Lambda - \frac{1}{4} F^2\right],
\end{equation}
where $F=dA$ is the $U(1)$ gauge field strength tensor, $R$ -- the Ricci scalar and $\Lambda = -3/L^2$ -- cosmological constant. 

As usual in holography \cite{Hartnoll:2009sz}, the chemical potential $\bm{\mu}$ and charge density $\bm{\rho}$ are related to the leading and subleading terms of the temporal gauge field component near the boundary of AdS (in what folows we denote the boundary field theory quantities with boldface): 
\begin{equation}
\label{equ:mu_rho}
A_t (z) \big|_{z\rar 0}  = \bm{\mu} + \bm{\rho} z.
\end{equation}
In the same fashion the radial current is related to the subleading term of the radial gauge field component 
\begin{equation}
\label{equ:Jr}
A_\rho(z) \big|_{z\rar 0}  = \bm{J}_\rho z.
\end{equation}
In what follows we will be studying the time- and space-resolved profiles of $\bm{\rho}(t,r)$ and $\bm{J}_\rho(t,r)$ when the local excitation in $\bm{\mu}(t,r)$ is created. 
Finally, the temperature of the field theory is controlled by the black hole horizon and is simply \cite{Witten:1998zw}:
\begin{equation}
\label{equ:temperature}
\bm{T} = \frac{3}{4 \pi z_h}.
\end{equation}
We will use the units of temperature in order to measure the dimensionful quantities.
% In what follows we will focus on the low temperature case
% \begin{equation}
% \label{equ:low_T}
% T/\mu = 0.01\,,
% \end{equation}
% which, together with \eqref{equ:temperature}, gives the following value of the chemical potential in units of horizon radius $z_h=1$:
% \begin{equation}
% \label{equ:mu}
% \mu \approx 3.22 z_h^{-1}\, .
% \end{equation}

\section{\label{sec:point_quench}Evolution after a local quench}
We are mostly interested in studying the non-equilibrium dynamics of the system \eqref{equ:RN} after a local quench in the profile of the chemical potential. We consider the local spherically-symetric perturbation with the Gaussian profile of the characteristic size $\rho_0$ which turns on at the moment $t=0$ and reaches the amplitude $\delta \mu$ at characteristic time $\tau_0$.
\begin{equation}
\label{equ:dmu}
\bm{\mu}(t,\rho) = - \delta \mu \cdot e^{- \frac{1}{2} \left( \frac{\rho}{\rho_0}\right)^2} \cdot \theta(t) \tanh\left(\tfrac{t}{\tau_0}\right).
\end{equation}
\begin{figure}[t]
\center
\includegraphics[width=0.45\linewidth]{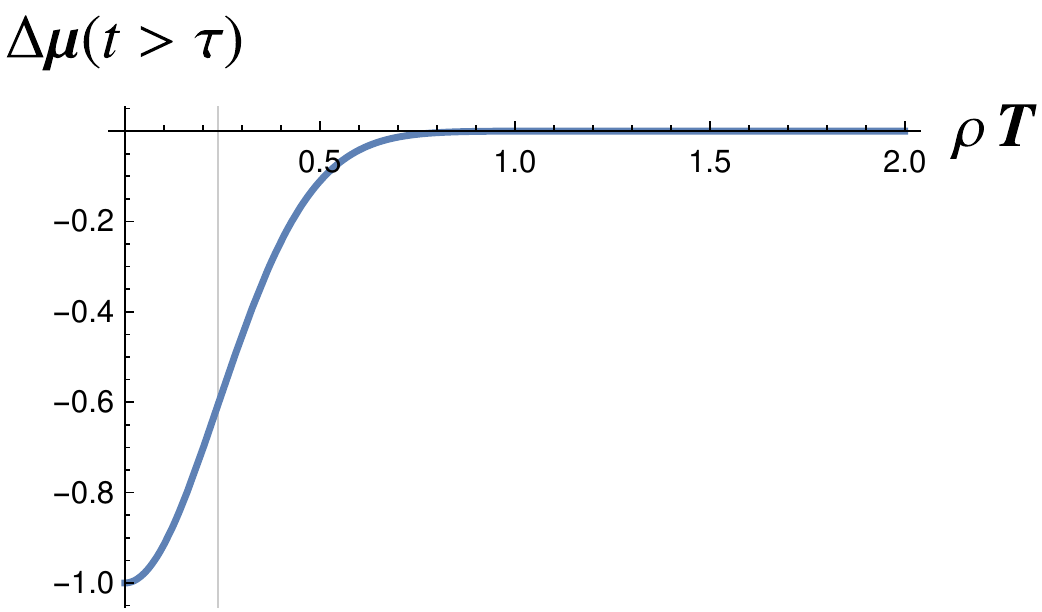}
\includegraphics[width=0.45\linewidth]{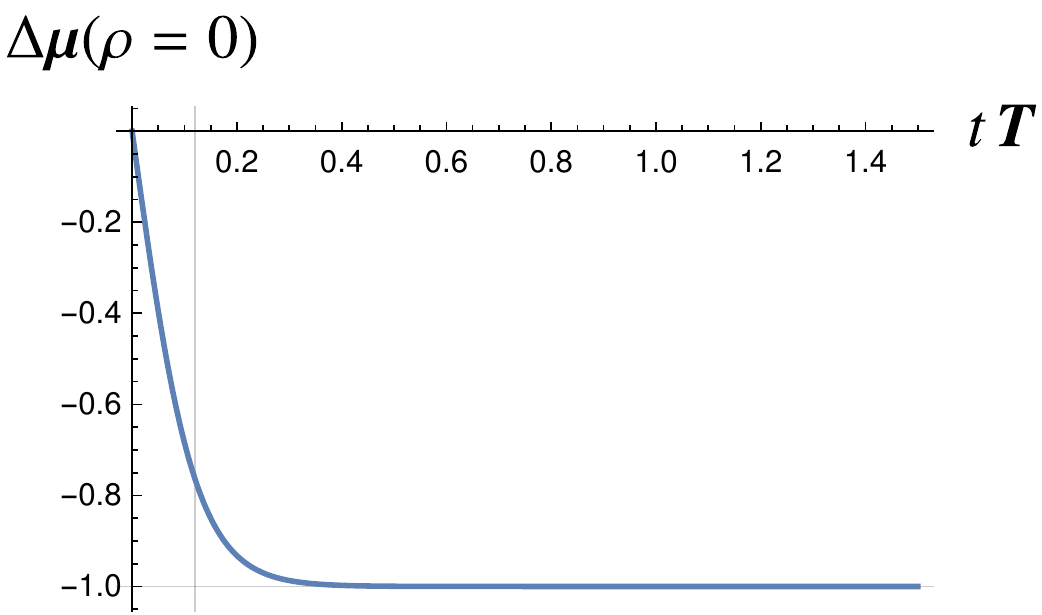}
\caption{\label{fig:dmuplot} \textbf{The profile of the perturbation in the chemical potential} \eqref{equ:dmu}. Left -- radial profile given the characteristic size $\rho_0=1 \, z_h = 0.25 \, \bm{T}^{-1}$ (shown by a gridline). Right -- temporal profile of the fast quench $\tau_0=0.5 \, z_h = 0.12 \, \bm{T}^{-1}$ (shown by a gridline).}
\end{figure}
In order to be able to treat this perturbation as local, but at the same time keeping it accessible to the numerical simulations, we chose the size to be reasonably small
\begin{equation}
\label{equ:size}
\rho_0 = 1 z_h \approx 0.25 \, \bm{T}^{-1} .
\end{equation}
We will keep $\tau_0$ as the tunable parameter and explore the behavior of the system in the different regimes when the quench is ``fast'', $\tau_0 \ll 1 \cdot z_h$ or ``slow'', $\tau_0 \gg 1 \cdot z_h$.

Our task therefore is to study the time-evolution of the system \eqref{equ:s_EM} with the gauge field subject to the boundary condition \eqref{equ:dmu} according to \eqref{equ:mu_rho}.
In order to describe the causal time-evolution we have to keep the infalling boundary condition at the black hole horizon. This is most straightforwardly done by choosing the generalized Eddington-Finkelshtein (EF) coordinates \cite{Chesler:2013lia}: $\{v,z,r,\phi\}$, where the time coordinate is replaced by
\begin{equation}
\label{equ:v_coord}
v = t - z^\ast, \qquad dz^\ast = dz f(z)^{-1}.
\end{equation}

In these coordinates the geodesics with constant $v$ correspond to the infalling waves, therefore the necessary horizon boundary condition reduces to the requirement of the regularity of solution in the EF coordinates \eqref{equ:v_coord}. The metric takes the form
\begin{gather}
\label{equ:EF}
ds^2 = \frac{1}{z^2} \left( - f(z) dv^2 - 2 dv dz + d\rho^2 + \rho^2 d\phi^2\right),
\end{gather}
Given that at $z\rar 0$ it follows from \eqref{equ:v_coord} that $z^\ast \sim z$, at the $AdS$ boundary the new coordinate $v$ equals boundary time-coordinate
\begin{equation}
\label{equ:v_t}
 v \big|_{z\rar 0} = t,
 \end{equation}
Therefore in the EF coordinates \eqref{equ:v_coord} one can directly use the boundary condition for $A_t$ \eqref{equ:dmu} replacing $t$ by $v$. The $z$-derivative of of fields depending on coordinate $v(t,z)$ is, however, modified (see Appendix \ref{app:EF}).

In this work we assume the amplitude of the perturbation to be small $\delta \mu \ll z_h^{-1}$. This allows us to linearize the equations of motion coming from \eqref{equ:s_EM} and neglect the backreaction of the gauge field to the metric. Let's study the spherically symmetric and time-dependent ansatz for the perturbed fields:
\begin{equation}\label{ansatz}
A = A_t(v,\rho,z) dt + A_\rho (v,\rho,z) d\rho + A_z(v,\rho,z)
\end{equation}
Keeping the arbitrary gauge we can recast the equations of motion in terms of the gauge invariant variables (these are just the components of fields strength $F_{\mu \nu}(t,\rho,z)$ in the original coordinates with extra normalization, which we will comment below)
\begin{gather}
\label{equ:definitions_Er}
\mathcal{F}_{\rho t} \equiv \p_\rho A_t - \p_v A_\rho, \qquad  \mathcal{F}_{z t} \equiv \p_z A_t - f(z)^{-1} \p_v A_t - \p_v A_z,\\
\label{equ:definitions_Ez}
 \mathcal{F}_{z \rho} \equiv \frac{1}{f(z)}\left(\p_z A_\rho - f(z)^{-1} \p_v A_\rho - \p_\rho A_z \right)
\end{gather}
Note that the identity holds
\begin{equation}
\label{equ:identity}
	\p_v \p_\rho \mathcal{F}_{z t} = f(z)^{-1} \p_v^2 \mathcal{F}_{z \rho} + \left(\p_z - f(z)^{-1} \p_v\right) \p_v \mathcal{F}_{\rho t}.
\end{equation}
The equations of motion on the background of the metric \eqref{equ:RN} in EF coordinates read
\begin{gather}
\label{EOM1}
f(z)\left(\p_z - f(z)^{-1} \p_v\right) \mathcal{F}_{z t} + \left(\p_\rho + \frac{1}{\rho}\right) \mathcal{F}_{\rho t} =0 \\
\label{EOM2}
f(z)\left(\p_z - f(z)^{-1} \p_v\right) \mathcal{F}_{z \rho} + \p_v \mathcal{F}_{\rho t} = 0\\
\label{EOM3}
\left(\p_\rho + \frac{1}{\rho}\right) \mathcal{F}_{z \rho} - \p_v \mathcal{F}_{z t} = 0 
\end{gather}
The $\p_v$ derivative of the \eqref{EOM1} is a linear combination of the derivatives of the other two equations, so provided \eqref{EOM1} is satisfied at $t=0$, it is enough for us to solve \eqref{EOM2} and \eqref{EOM3}.

We can act with $\left(\p_z - f(z)^{-1} \p_v\right)$ on \eqref{EOM2} and with $\p_\rho$ on \eqref{EOM3}. Then after taking into account the identity \eqref{equ:identity}, we can eliminate the second derivatives in $v$. This allows us to derive a single evolution equation on $\mathcal{F}_{z \rho}$.
\begin{equation}
\label{equ:EOMs}
\left( \p_\rho^2 + \frac{1}{\rho} \p_\rho - \frac{1}{\rho^2}\right) \mathcal{F}_{z \rho} + f(z) \p_z^2 \mathcal{F}_{z \rho} + f'(z) \p_z \mathcal{F}_{z \rho} - 2 \p_v \p_z \mathcal{F}_{z \rho} = 0 
\end{equation}

In what follows we solve the evolution equation \eqref{equ:EOMs} subject to the time-dependent boundary condition for $A_t$, which is set by the perturbation of the chemical potential \eqref{equ:dmu}. First of all, we figure out how this boundary condition translates to $\mathcal{F}_{z\rho}$.  
For this purpose it is most convenient to use equation \eqref{EOM2}, which, given the definition of $\mathcal{F}_{\rho t}$ \eqref{equ:definitions_Er} reduces to the Robin boundary condition on $\mathcal{F}_{z \rho}$ at at $z=0$
\begin{equation}
\label{equ:BCB}
z=0: \qquad \left(\p_z - \p_v \right) \mathcal{F}_{z \rho} = - \p_v \p_\rho \bm{\mu}(v,\rho)
\end{equation}
The boundary conditions at $\rho=0$ and $\rho \rar \infty$ are set by the generic symmetry structure of the polar coordinates and the requirement that the solution should approach the profile of the neutral Schwarzschild black hole \eqref{equ:RN} at infinity
\begin{equation}
\label{equ:BCL}
\rho=0: \qquad  \mathcal{F}_{z\rho}(v,0,z)  = 0,  \qquad \rho \rar \infty: \qquad \mathcal{F}_{z\rho}(v,\rho,z)  \rar 0.
\end{equation}
In the end of the day, as discussed above, we require regularity of solution at the horizon. The asymptotic analysis of the master equation \eqref{equ:EOMs} at $z=1$ shows that, thanks to the extra factor of $f(z)$ introduced in \eqref{equ:definitions_Ez}, the function $\mathcal{F}_{z\rho}(v,0,z)$ is finite and satisfies the Robin boundary condition in $z$:
\begin{align}
\notag
z = z_h: \qquad & \mathcal{F}_{z\rho}(v,\rho,1)  = \mathit{finite}, \\
\label{equ:BCT}
 & 2 \, \p_v \p_z \mathcal{F}_{z\rho} + 3 \, \p_z \mathcal{F}_{z\rho} = \left(\p_\rho^2  + \frac{1}{\rho} \p_\rho + \frac{1}{\rho^2} \right) \mathcal{F}_{z\rho}.
\end{align}

The evolution equation \eqref{equ:EOMs} with the boundary conditions \eqref{equ:BCB},\eqref{equ:BCL},\eqref{equ:BCT} can be solved numerically by applying the ``method of lines'' (see Appendix \ref{app:lines}).

Once we obtain the solution at all times, we can extract the time-dependent values of the observables using the asymptotic formulae \eqref{equ:mu_rho} and \eqref{equ:Jr}. Note that since the time coordinates $v$ and $t$ are identical at the boundary, we directly get the observables in terms of~$t$. The appropriate $z$-derivatives are already incorporated in the gauge invariant functions $\mathcal{F}_{z\rho}$ and $\mathcal{F}_{z t}$ \eqref{equ:definitions_Ez},\eqref{equ:definitions_Er}, which we are solving for: the value of the observable current is exactly the boundary value of the field $\mathcal{F}_{z\rho}$:
\begin{equation}
\label{equ:current}
\bm{J}_\rho (t,\rho) = \mathcal{F}_{z\rho}(v,\rho,z) \big|_{v=t, z=0}
\end{equation}
Similarly the value of the observable charge density is $\bm{\rho} (t,\rho) = \mathcal{F}_{zt}(v,\rho,z) \big|_{v=t, z=0}$. It can be obtained from the equation \eqref{EOM3}, which on the boundary simply turns into the continuity equation:
\begin{equation}
\label{equ:charge}
\p_t \bm{\rho} (t,\rho)  = \left(\p_\rho + \frac{1}{\rho}\right) \bm{J}_\rho(t,\rho).
\end{equation}
In what follows we obtain the value of the charge density by integrating \eqref{equ:charge} along the whole evolution.

\begin{figure}[t]
\center
\begin{tabular}{cc}
\begin{minipage}{0.5 \linewidth}
\center
\includegraphics[width=1\linewidth]{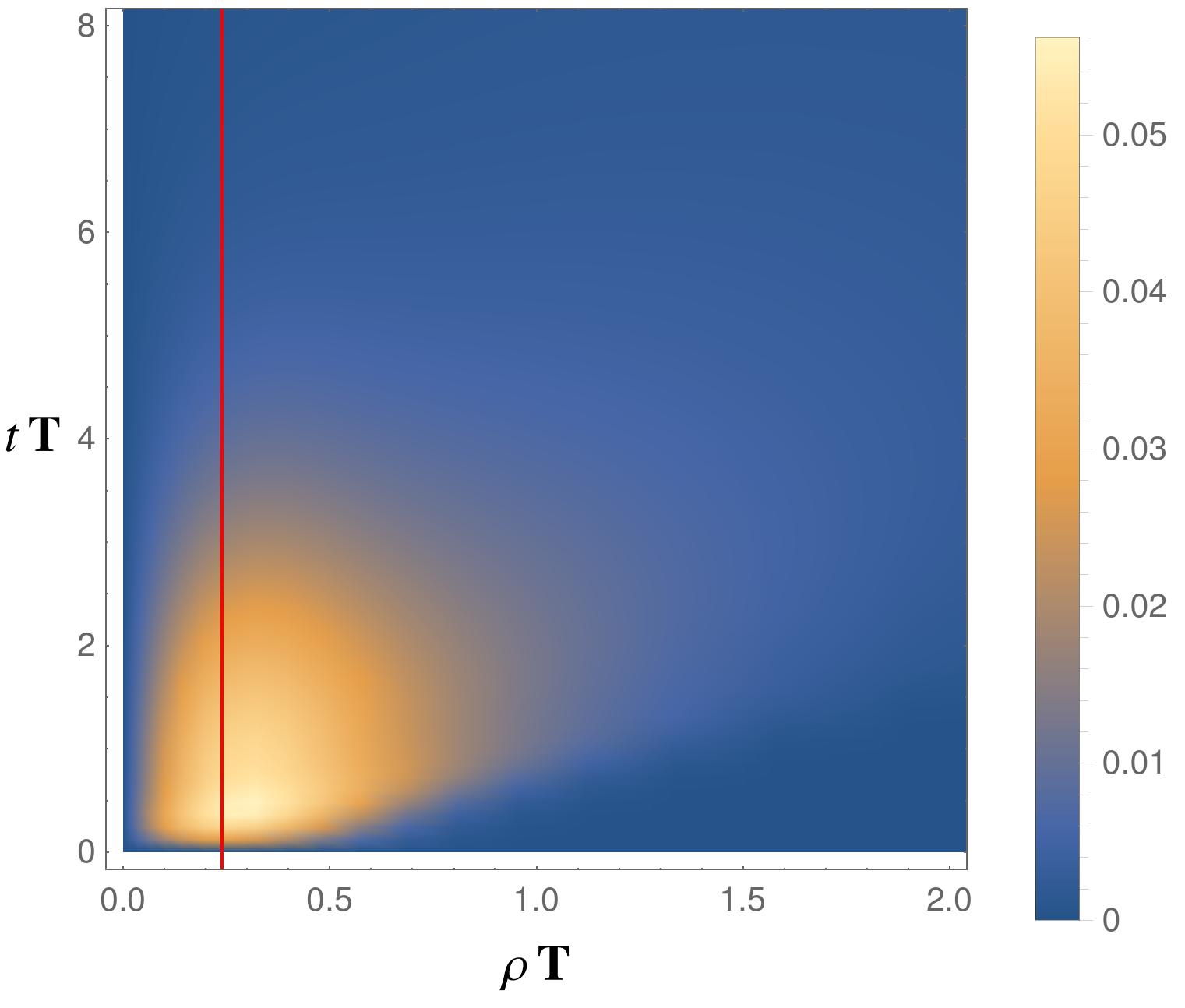}
\end{minipage}
&
\begin{minipage}{0.45 \linewidth}
\center
\includegraphics[width=1\linewidth]{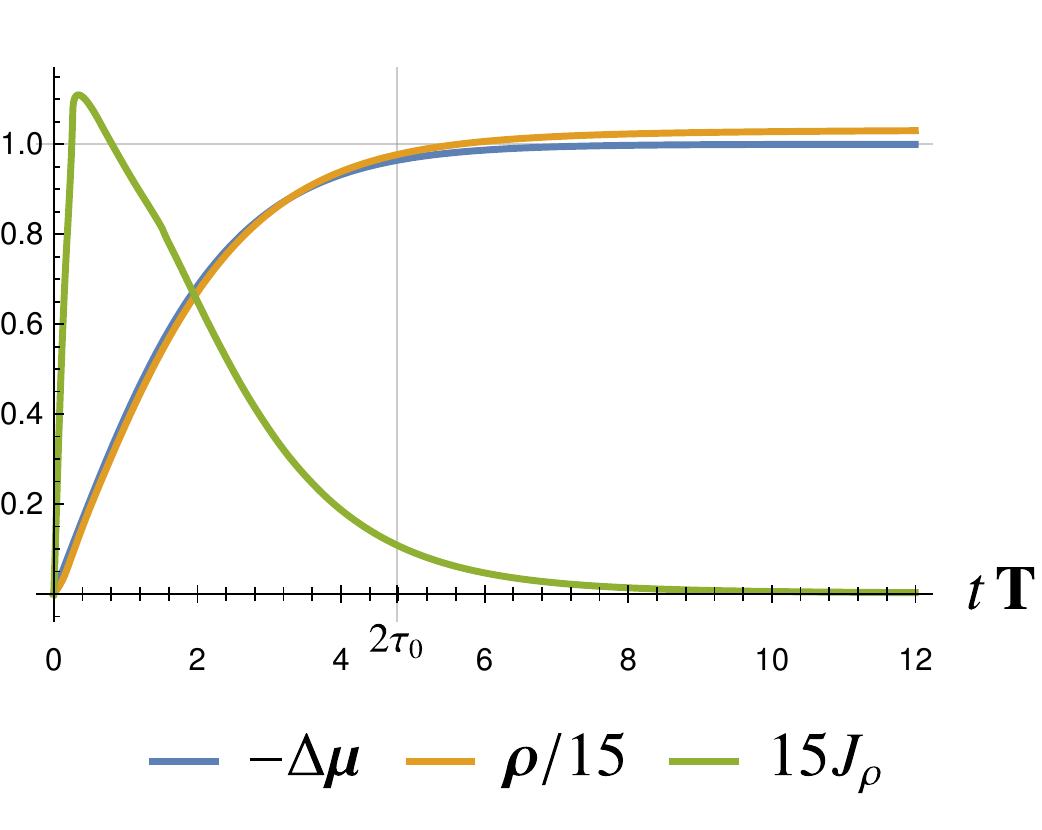}
\end{minipage}
\end{tabular}

\caption{\label{fig:slowQ} \textbf{Slow quench} $\tau_0=10 z_h \approx 2.4 \bm{T}^{-1}$. Left: the profile of the radial current depending on time $t$. No propagating waves are seen. Right: the time-dependence of the local chemical potential and charge density at $\rho=0$ and the current density at $\rho \sim \rho_0 \approx 0.25 \bm{T}^{-1}$ (characteristic radius of the initial quench, shown by red gridline on the left).}
\end{figure}

If we consider the \textbf{slow rise of the perturbation potential}, ensuring that its characteristic timescale $\tau$ is long enough for the system to assume the local equilibrium at any given moment, we expect to observe the adiabatic evolution. Indeed, in case of the slow quench $\tau=10 z_h = 2.4 \, \bm{T}^{-1}$ we see that all the observables change with the same rate as the quench itself, see Fig.\,\ref{fig:slowQ}. And the timescale is set by $\tau_0$. The system undergoes a sequence of quasi-equilibrium states with the constant ratio between the charge density and the chemical potential (blue and yellow curves on the right panel of Fig.\,\ref{fig:slowQ}). The slow quench truly realizes the adiabatic process.

\begin{figure}[t]
\center
\begin{tabular}{cc}
\begin{minipage}{0.5 \linewidth}
\center
\includegraphics[width=1\linewidth]{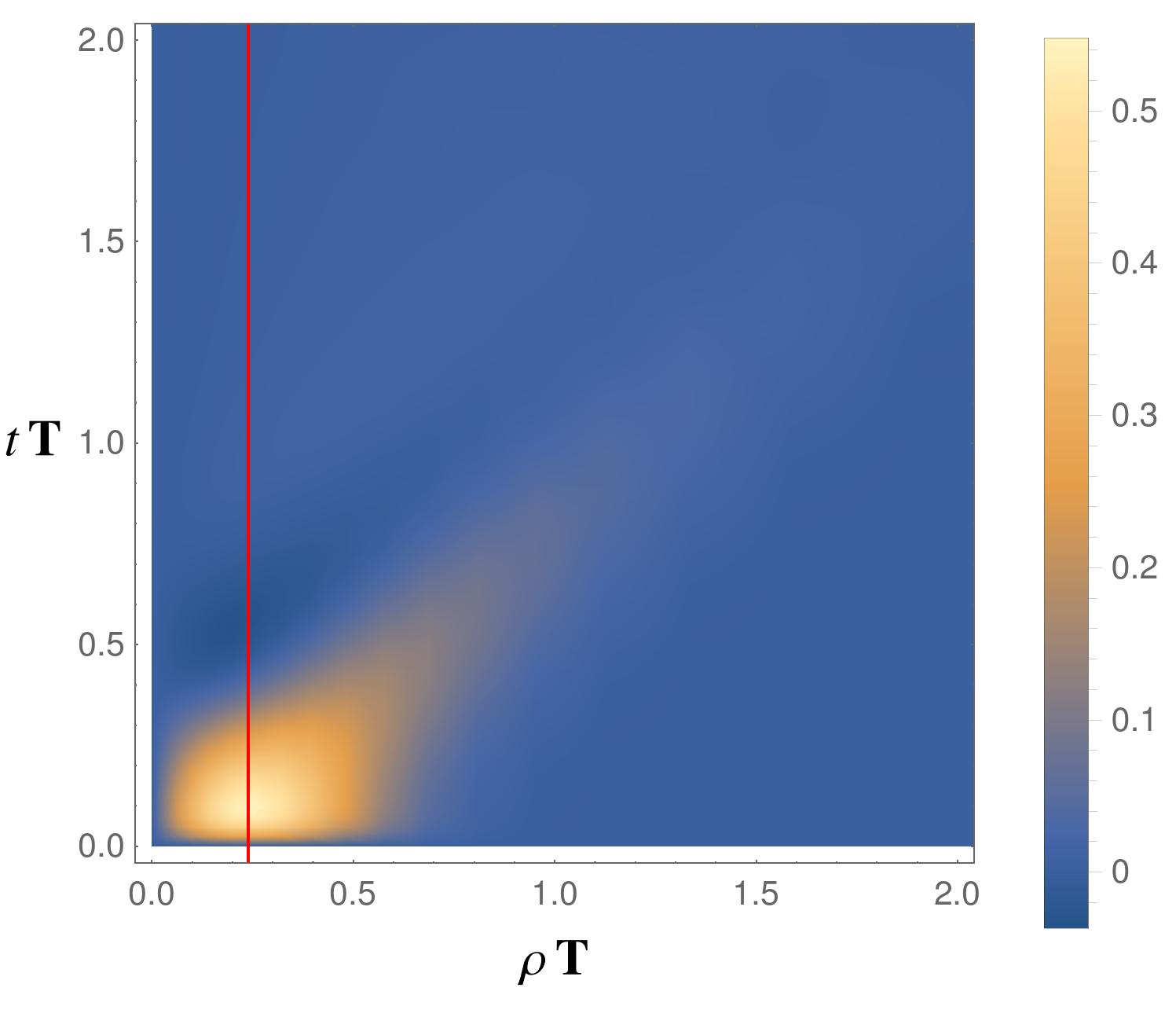}
\end{minipage}
&
\begin{minipage}{0.45 \linewidth}
\center
\includegraphics[width=1\linewidth]{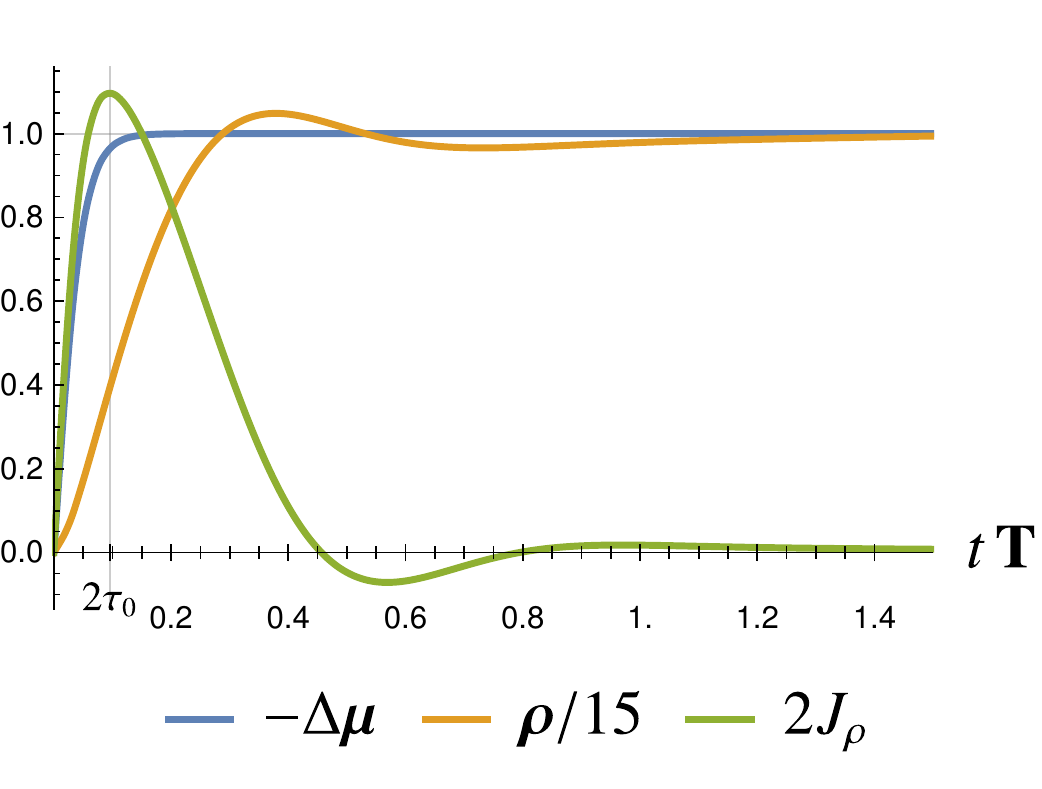}
\end{minipage}
\end{tabular}

\caption{\label{fig:fastQ} \textbf{Fast quench} $\tau= 0.5 z_h = 0.12 \bm{T}^{-1}$. Left: the profile of the radial current depending on time $t$. The expanding wave of the current at $t \ll \tau$ is visible. Right: the time-dependence of the local chemical potential and charge density at $\rho=0$ and the current density at $\rho \approx 0.25 \bm{T}^{-1}$ (shown by red gridline on the left)}
\end{figure}

In case of the \textbf{fast quench} ($\tau_0 = 0.5 z_h = 0.12 \, \bm{T}^{-1}$), shown on Fig.\,\ref{fig:fastQ}, the situation is substantially different. Now the timescale of the relaxation of the system is much larger then $\tau_0$ and is set by the internal dynamics. The charge density now lags behind the evolution of the chemical potential (see the time profiles on the right panel of Fig.\,\ref{fig:fastQ}), while eventually arriving to the same static state at late times.\footnote{The match between the end values of the charge density in cases of slow and fast quenches serves as a good check of our numerical procedure, since the value of charge density is obtained as an integral over the whole evolution \eqref{equ:charge}} Note also that the radial current profile is order of magnitude larger then that in case of the slow quench. The spatial spread of the excitation is also going beyond the size of the initial quench: the expanding current density wave is visible on the left panel of Fig.\,\ref{fig:fastQ}, which was absent in the case of the slow quench.
All these features point out that the system is far from equilibrium and the relaxation process is governed by the internal dynamics, independent of the precise profile of the quench. 

\section{\label{sec:QNMs}Spherical quasinormal modes}

The internal time dynamics of the system, which governs the relaxation after the fast quench on Fig.\,\ref{fig:fastQ}, can be studied by means of the quasinormal modes (QNM) spectrum analysis. The QNMs have finite negative imaginary part due to the absorption by the horizon, which is controlled by the temperature of the system. The spectrum of QNM can be obtained as a solution to the Sturm-Liouville problem for the equations of motion with trivial boundary condition for the oscillatory modes of the form $A_\mu \sim e^{- i \omega t}$.
The values of $\omega$, for which the equations have a nontrivial solution give the spectrum of QNMs \cite{Kovtun:2005ev}. 

When studying the oscillatory modes we can take advantage of the cylindrical symmetry of the problem and express the solution in terms of the Bessel functions. In other words, we are looking for the spectrum of \textit{cylindrical} quasinormal modes:
\begin{align}
\label{equ:QNM}
A_t(t,\rho,z) &= a_t(z) e^{- i \omega t} \mathcal{J}_0 (\alpha \rho), \\
\notag
A_\rho (t,\rho,z) &= a_\rho(z) (-i)  e^{- i \omega t} \mathcal{J}_1 (\alpha \rho), \\
\notag
A_z(t,\rho,z) &= a_z(z)  e^{- i \omega t} \mathcal{J}_0 (\alpha \rho),
\end{align}
where the spatial scale $\alpha$ plays the role analogous to the wave-vector in plain wave excitations. In this ansatz the radial and time dynamics of the problem decouple completely and we are left with a system of ODEs:
\begin{gather}
\label{equ:ODE_QNMs}
\p_z^2 a_t - \frac{\alpha^2}{f(z)} a_t + i \omega \p_z a_z + \frac{\alpha \omega}{f(z)} a_\rho = 0, \\
\notag
\p_z^2 a_\rho + \frac{f'(z)}{f(z)} \p_z a_\rho + \frac{\omega^2}{f(z)^2} a_\rho - \frac{\alpha \omega}{f(z)^2} a_t + i \alpha \p_z a_z + \frac{i \alpha f'(z)}{f(z)} a_z = 0, \\
\notag
\left(\frac{\omega^2}{f(z)} - \alpha^2 \right) a_z + i \alpha \p_z a_\rho - \frac{i \omega}{f(z)} \p_z a_t = 0.
\end{gather}
Again, we can use the gauge invariant variables to further simplify the problem \cite{Kovtun:2005ev}:
\begin{equation}
e_\rho \equiv - \omega a_\rho + i \alpha a_t, \qquad e_z = - i \omega a_z - \p_z a_t.
\end{equation}
The system \eqref{equ:ODE_QNMs} is therefore rewritten in terms of only two functions and decouples. We end up with
\begin{align}
\label{equ:gauge_invar_QNMs}
\p_z^2 e_\rho + \frac{\omega^2}{\omega^2 - \alpha^2 f(z)} \frac{f'(z)}{f(z)} \p_z e_\rho + \frac{\omega^2 - \alpha^2 f(z)}{f(z)^2} e_\rho = 0 \\
\notag
\left(\alpha^2 f(z) - \omega^2 \right) e_z + \alpha f(z) \p_z e_\rho = 0.
\end{align}
The first equation here is a master equation and we can extract the spectrum of QNMs from there. Noteworthy, this equation is exactly the same as the one, which defines the spectrum of plain wave QNMs of the 4D-AdS-Schwarschild background \cite{Miranda:2008vb}. It is convenient to setup the boundary conditions at $z\rar0$ directly on $e_\rho$, since this object is related to the radial component of the electric field and is completely set by the boundary data. Therefore we simply look for solutions to \eqref{equ:gauge_invar_QNMs} with
\begin{equation}
\label{equ:BCB_QNM}
e_\rho |_{z\rar0} =0.
\end{equation}
On the other hand, the analysis near horizon reveals that the solution is a superposition of the infalling and outgoing waves, of which we take only the infalling component (note that unlike previous Section, we are not working in EF coordinates here):
\begin{equation}
\label{equ:BCH_QNM}
e_\rho |_{z\rar z_h} \sim \left(1-\tfrac{z}{z_h} \right)^{- i \omega z_h/3}.
%, \left(1-\tfrac{z}{z_h} \right)^{- i \omega/3}
\end{equation}
%We pick the exponent with minus sign, since in conjunction with $e^{- i \omega t}$ it forms an infalling wave under horizon.
The Sturm-Liouville problem  for the ordinary differential equation \eqref{equ:gauge_invar_QNMs} subject to the boundary conditions \eqref{equ:BCB_QNM} and \eqref{equ:BCH_QNM} can be easily analyzed by means of numerical shooting method, as we describe in more details in Appendix \ref{app:QNMs}. 

\begin{figure}[t]
\center
 \raisebox{-0.5\height}{\includegraphics[width=0.58\linewidth]{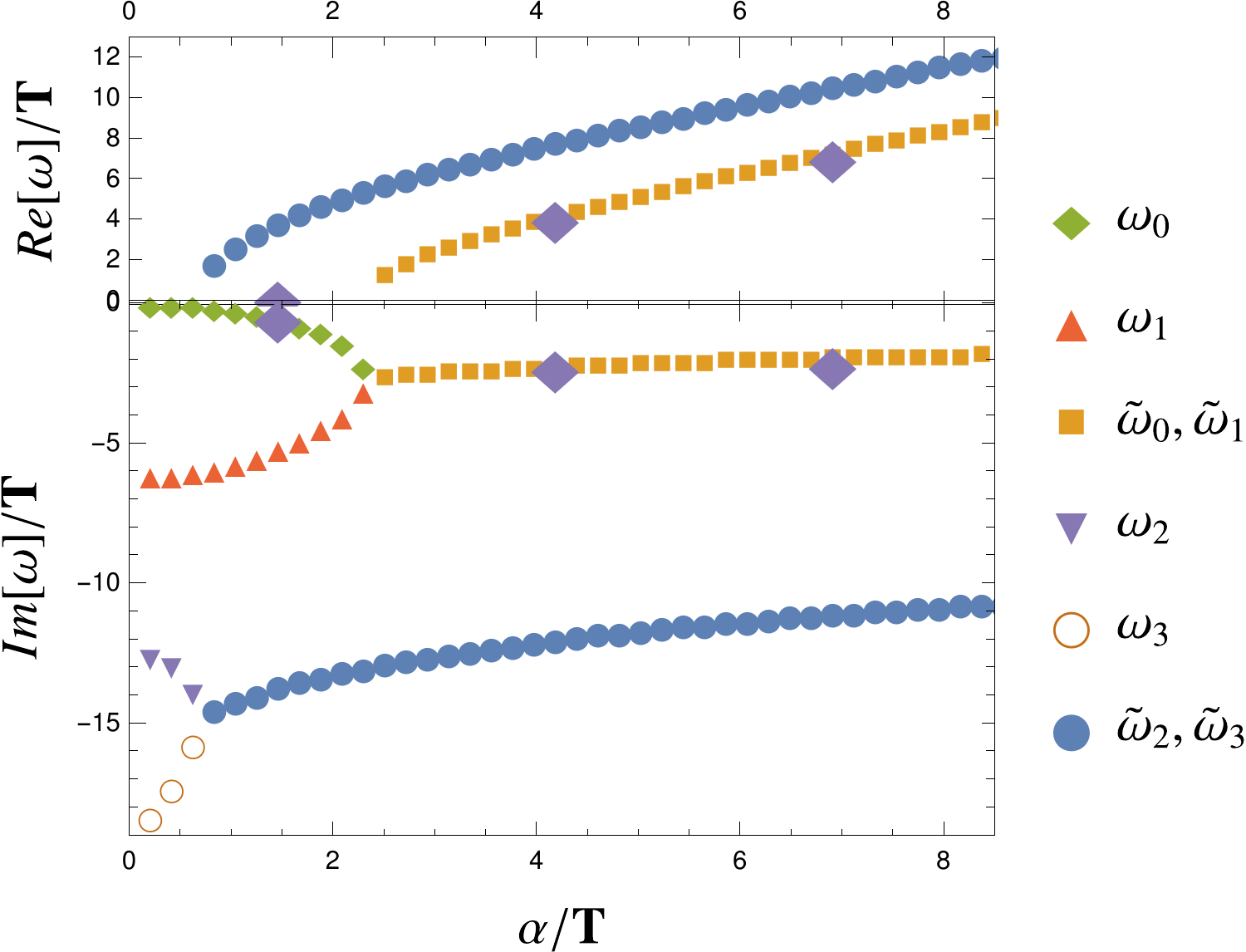}}
 \raisebox{-0.5\height}{\includegraphics[width=0.4\linewidth]{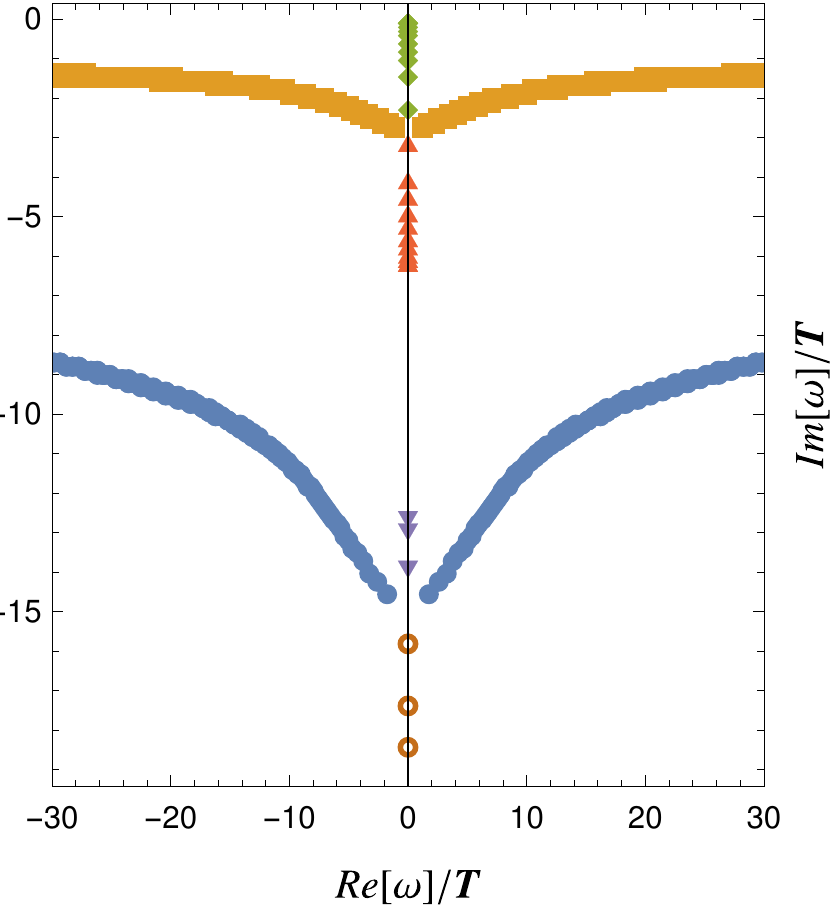}}
\caption{\label{fig:QNMplot} \textbf{Cylindrical quasinormal modes} of the Schwarzschild black hole \eqref{equ:RN}. Left panel: real and imaginary parts as functions of the radial scale parameter $\alpha$, \eqref{equ:QNM}. Right panel: QNM positions on the complex $\omega$ plane for all radial scale parameters $\alpha \in (0,32)\, \bm{T}$ overlaid. The evolution of 4 modes is visible, going from purely decaying $Re(\omega_i) = 0$ to oscillatory pairs $Re(\tilde{\omega}_i) \neq 0$. The mode $\omega_0$ can be identified with the hydrodynamic charge diffusion, while the others are non-hydrodynamical. Purple diamonds on the left panel show the parameters of the modes extracted from late time analysis, see Table \ref{QNMfit}.}
\end{figure}

For a given ``wave-vector'' $\alpha$ we obtain a set of values of $\omega_i(\alpha)$ in the imaginary plane, for which the nontrivial solution exists. These are shown on Fig.\,\ref{fig:QNMplot}. On the left panel we show the evolution of the QNMs as $\alpha$ is increased up to $8\,\bm{T}$. In our scope we observe the 4 purely imaginary modes, in ``long-wavelength'' regime, which collide pairwise and form 2 pairs of oscillatory modes with finite real parts when $\alpha$ grows. This is exactly the same behavior as the one observed for the plain-wave polar electromagnetic QNMs in this background in \cite{Miranda:2008vb}. The mode $\omega_0$, shown in green diamonds is associated to the diffusion of the electric charge in the system and can be obtained from the hydrodynamic analysis. However the higher modes are specific to the black hole and do not admit the hydrodynamic description. 

When the system undergoes a quench, the quasinormal modes are excited and further relaxation is controlled by the decay of these modes (with the timescale set by their imaginary part). In case of the Gaussian spatial profile of the quench, the QNMs with different $\alpha$'s are excited, according to the spectral representation of the Gaussian with Bessel functions, the more localized is the quench -- the more modes are included. The distribution of all possible excited modes is shown on the right panel of Fig.\,\ref{fig:QNMplot}. We see two branches of the oscillatory modes, with imaginary positions in intervals $Im(\omega) \in (-3\,\bm{T},-1\,\bm{T})$ for $\tilde{\omega}_0,\tilde{\omega}_1$ and $Im(\omega) \in (-15\,\bm{T},-8\,\bm{T})$ for $\tilde{\omega}_2,\tilde{\omega}_3$, for a wide range of the radial scale parameter $\alpha \in (0, 32\,\bm{T})$. One can see, that apart from the diffusion mode (green), the slowest decay rate of the oscillating modes (yellow) saturates at around $\approx 2. \bm{T}$. Therefore one expects most of the excited modes to decay at this rate and indeed this agrees with the observed decay time of the peak on Fig.\,\ref{fig:fastQ} is 
\begin{equation}
\tau_{\mathrm{main\,peak\,decay}} \approx 0.5 \, \bm{T}^{-1} \approx Im[\tilde{\omega}_1]^{-1}.  
\end{equation}
The structure of the excited quasinormal modes also controls the expanding spread of the current wave, which is seen for the fast quench: the spatial structure of the response, like the temporal one, is not defined by the features of the quench, but rather by the radial scale parameter $\alpha$ of the excited quasinormal mode.

\begin{figure}[t]
\center
\includegraphics[width=0.49\linewidth]{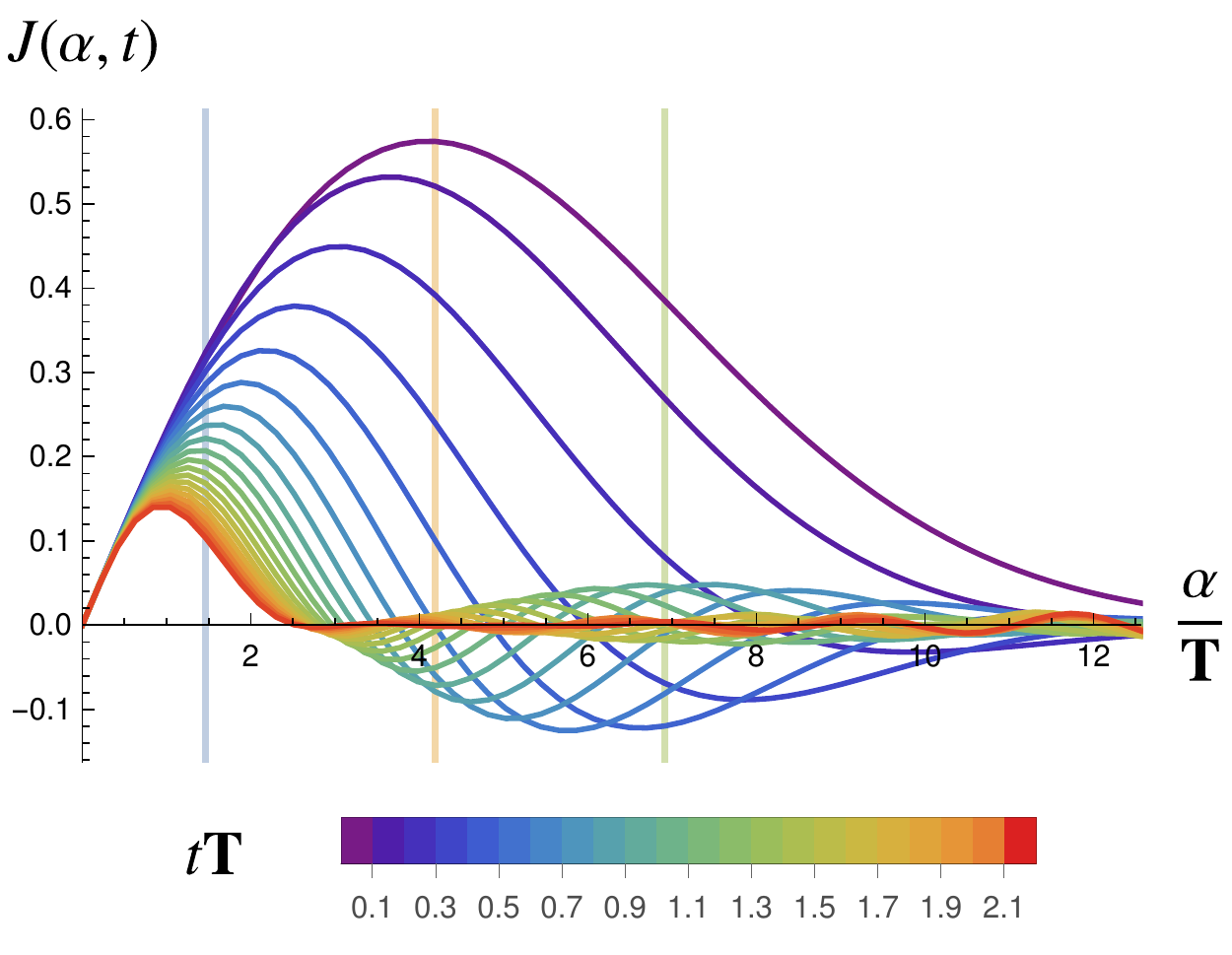}
\includegraphics[width=0.49\linewidth]{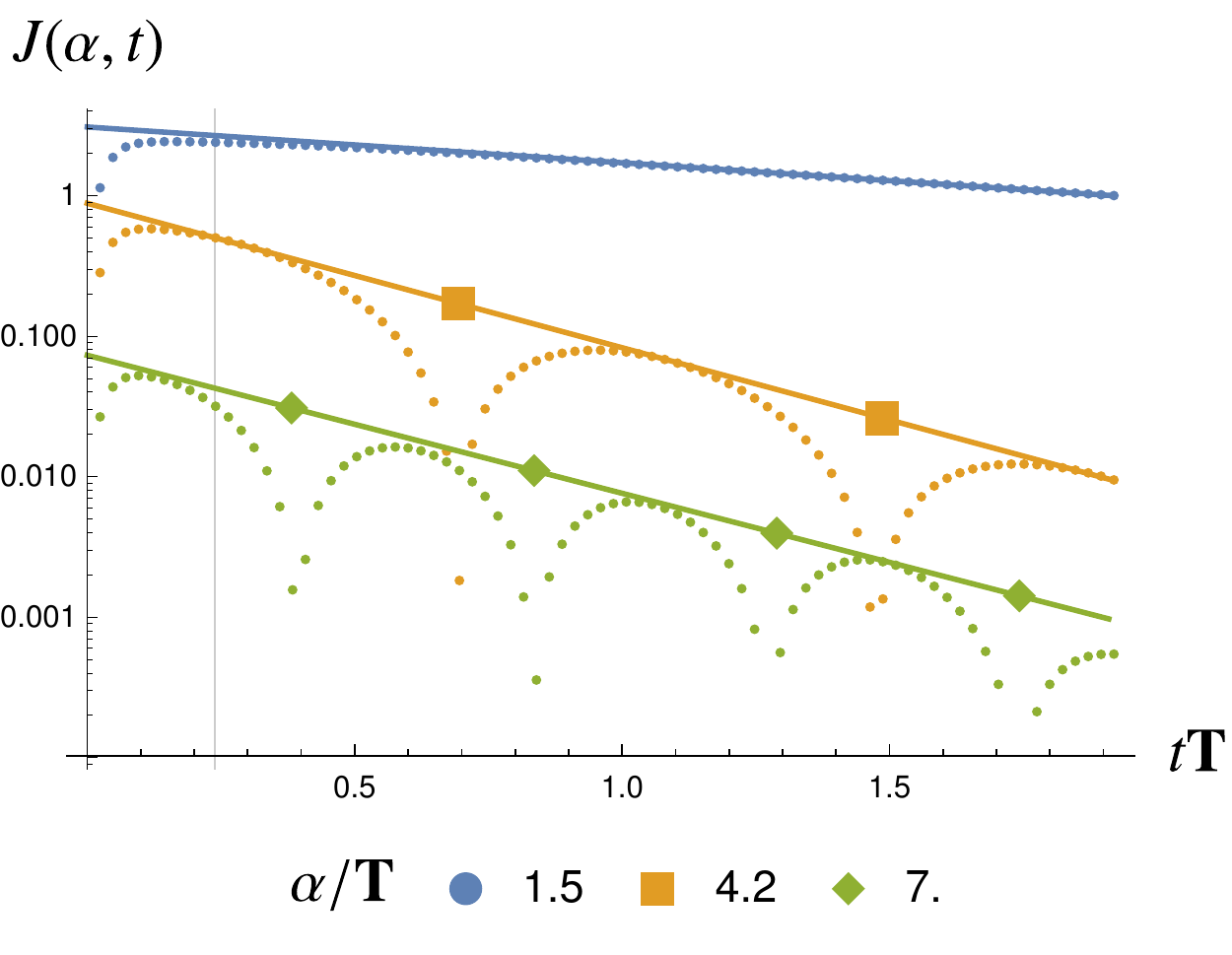}
\caption{\label{fig:late_time} \textbf{Late time relaxation analysis} after the fast quench. \\
\textbf{\textit{Left panel:} time evolution of the Bessel spectrum}. The modes with $\alpha>2$ decay faster and oscillate, while the modes with $\alpha<2$ decay slower in pure exponential fashion. This is in qualitative agreement with the spectrum of QNMs, Fig.~\ref{fig:QNMplot} \\
\textbf{\textit{Right panel:} decay of the selected Bessel modes}, shown by grid lines on the left panel. The exponential slope and periodic oscillations are clearly visible on the logarithmic scale. The fits listed in Table \ref{QNMfit} agree quantitatively with the  the spectrum of QNMs, Fig.~\ref{fig:QNMplot}.}
\end{figure}

We can capture the contribution of the quasinormal modes in the late time decay of the signal more precisely by expanding the dynamical evolution after the fast quench of the previous section in a spectrum of Bessel functions. In particular, at every given moment $t$ we perform a Hankel transform of the spatial profile of the current (note that according to \eqref{equ:QNM} the expansion the expansion goes in the Bessel functions of rank 1):
\begin{equation}
\bm{J}(\alpha, t) \equiv \int \limits_0^\infty d \rho \rho \mathcal{J}_1 (\rho \alpha) \, \bm{J}(\rho, t).
\end{equation}
The resulting spectrum is depicted on the left panel of Fig.\,\ref{fig:late_time}. First of all, we see that the initial spectrum is precisely the one associated with the Gaussian quench profile \eqref{equ:dmu} with maximum at $\alpha = 1/\rho_0 \approx 4 \, \bm{T}$. We also see that the modes with $\alpha>2 \, \bm{T}$ decay relatively quickly and oscillate around zero, as expected from the behavior of oscillating QNMs $\tilde{\omega}_0,\tilde{\omega}_1$. However at $\alpha < 2 \, \bm{T}$ the decay is much slower and no oscillations are present: this is the part of the spectrum dominated by diffusion mode $\omega_0$. This result is in perfect qualitative agreement with the spectrum of Fig.\,\ref{fig:QNMplot}.

From the late time fall off of these Bessel spectral modes we can actually independently extract the parameters of the lowest lying QNMs, which provides an excellent cross-check of our analysis. Indeed, the evolution of the selected set of the Bessel modes with $\alpha = \{1.5, 4.2, 7\} \, \bm{T}$ is shown on the right panel of Fig.~\ref{fig:late_time} in logarithmic scale. One can clearly see the exponential fall off of the diffusion mode at $\alpha=1.5 \, \bm{T}$ and the exponentially decaying oscillations for the higher Bessel harmonics. From the slopes and the period of zeros of these curves we can extract both real and imaginary parts of the corresponding QNMs, see Table \ref{QNMfit}, which turn out to be in excellent agreement with the computed QNM spectrum, see purple diamonds on the left panel of Fig.\ref{fig:QNMplot}. 
\begin{table}[t]
\centering
\begin{tabular}{|l|l|l|l|}
\hline
Bessel mode ($\alpha$) & Exponential slope & Half-Period & extracted QNM  \\
\hline
1.46 & -0.586 & $\infty$ & 0.00 - I 0.59 \\
4.19 & -2.37 & 0.79 & 3.96 - I 2.37 \\
6.91 & -2.26 & 0.45 & 6.92 - I 2.26 \\
\hline
\end{tabular}
\caption{\label{QNMfit} The parameters of the QNMs extracted from the late time analysis of the Bessel spectrum after the fast quench. The imaginary part of the QNM is measured from the exponential slope visible on the log-scale plot, Fig.~\ref{fig:late_time}, right panel. The real part is evaluated from the half period of oscillations $Re[\omega]= \frac{\pi}{\lambda/2}$. All values in units of temperature $\bm{T}$.}
\end{table}

In the end of the day the result of our analysis shows that the relaxation after the fast local quench is indeed very well captured by the spherical quasinormal modes.
Importantly, according to \eqref{equ:QNM}, the relaxation of these modes is always exponential. The modes with finite real part correspond to the exponentially decaying oscillations. The quasinormal modes do not lead to the power law time tails, which would be expected for the conventional orthogonality catastrophe-like behavior.

\section{\label{sec:concl}Conclusion}

In this work we consider the non-equilibrium evolution of the charge symmetric (chemical potential is zero) system of strongly correlated electrons after the local change in the chemical potential. This is a model problem for the X-ray spectrometry experiment, which in case of the weakly coupled electrons demonstrates the interesting quantum phenomenon of the X-ray edge anomaly. Our charge symmetric setup may correspond to the dynamics of graphene at half filling, which can be understood as a strongly correlated liquid of emergent relativistic degrees of freedom. We use the holographic duality to describe the system and we work in the probe approximation, neglecting the backreaction of the gauge field profile on the geometry.

In the simple setup under consideration we observe that the dynamics of the system, even far from equilibrium, is described by the linear equation of motion \eqref{equ:EOMs}. In this case the spectrum of quasinormal modes gives an exact representation of the possible non-equilibrium solutions. The internal time-scale of the QNM relaxation $\tau_{\mathrm{decay}}$, which is set by the temperature, allows one to distinguish between the two characteristic types of the evolution after a local quench. If the quench is slow and the defect develops at the time-scales longer then the equilibration time $\tau_{\mathrm{decay}}$, the dynamics is adiabatic. On the other hand, when the quench is fast, the QNMs are excited and the relaxation dynamics has an oscillatory behavior with expanding waves of current.

In the present study we develop several techniques and approaches which will be useful in the future investigations: the evolution equation in terms of gauge invariant quantities, the method of lines for the integration of the gauge field evolution equations and the spherically symmetric setup suitable for treatment of the local defects and corresponding QNMs.

We do not find a regime with the power law relaxation pattern, which would be similar to the pattern of the orthogonality catastrophe. This finding, although its nature its quite easy to understand from the holographic standpoint, may serve as an indirect evidence for the exponential orthogonality catastrophe in the charge neutral graphene and many body localized systems, studied previously in \cite{yang2007x,deng2015exponential}. On the other hand, it certainly calls for future investigations, including the systems at finite charge density. The obvious next step is to consider the Reissner-Nordstr\"om background and include the coupling of the perturbations to the metric fields. In this setup the zero temperature limit can also be studied, where the branch cuts may appear in the QNM spectrum, breaking down the exponential behavior. Further on, inclusion of the full backreaction will make the evolution equations nonlinear and will allow for the solutions which are not representable by the linear combination of the QNMs. The other possible generalization is to include the features of the finite $N$ by making use of the Gauss-Bonet gravitational setups.

This work serves as a proof of principle that the holographic duality provides an adequate tool for treating the time-evolution problems in the system of strongly correlated electrons and has its advantages as compared to the adiabatic perturbation theory and the Schwinger-Keldysh approach. 

\acknowledgments{
I thank Oleg Lychkovkiy, Oleksandr Gamayun and Koenraad Schalm for proposing this problem. I benefited a lot from the useful discussions with Dmitry Ageev, Andrey Bagrov and Mikhail Katsnelson about the local quenches. I'm grateful to Aurelio Romero-Bermudez and Christian Ecker for discussing the time-evolution problems in holographic setups. I especially appreciate the advice of Julian Sonner, Ben Withers and Tomas Andrade, who provided me with many useful references on the subject.

The work was supported by the Russian Science Foundation under the grant \hbox{No. 17-71-20158}.
}

\appendix

\section{\label{app:lines} Method of lines}
The numerical ``method of lines'' allows one to solve the evolution differential equations with partial derivatives (PDEs), which describe the non-equilibrium dynamics of the system with a local defect \eqref{equ:EOMs}. One introduces the calculation grid $\{\rho_i, z_j\}$ in the spatial part of the integration region and represents the derivatives as finite differences of the functional values on this grid. After the discretization of the spatial directions the differential equation \eqref{equ:EOMs} is represented by the set of $\times N_\rho \times N_z$ ordinary differential equations with only dependence on the ``temporal'' coordinate $v$ on the functions
\begin{equation}
\mathcal{F}_{z \rho}^{ij} (v) \equiv \mathcal{F}_{z \rho}(v,\rho_i,z_j)
\end{equation}
and their derivatives in $v$. More concretely, the differential operators of \eqref{equ:EOMs} can now be represented as the matrix operators acting on the fields $\mathcal{F}_{z \rho}^{ij} (v)$ and their time derivatives
% \begin{multline}
%  \begin{pmatrix}
%  \left[- \frac{1}{f(z)}\p_z  + \frac{f'(z)}{f(z)^2} \right] & 0  \\
% \frac{1}{f(z)}  \p_\rho & - 2  \p_z 
% \end{pmatrix}
% \begin{pmatrix}
%  \p_v A_t \\  \p_v A_\rho
%  \end{pmatrix} 
%  + \\
% \begin{pmatrix}
%  \left[ \p_z^2 + \frac{1}{f(z)}\p_\rho^2 + \frac{1}{r f(z)} \p_\rho \right]&  - \left[\frac{1}{\rho}\p_z + \p_\rho \p_z \right]  \\
%  0 & \left[ f(z) \p_z^2 + f'(z) \p_z \right] 
% \end{pmatrix}
% \begin{pmatrix}
%  A_t \\  A_\rho
%  \end{pmatrix} =0 
% \end{multline}
%
%In a more concise notation one can write it down as
\begin{equation}
\label{equ:evolution}
O_1 \p_v \vec{\mathcal{F}_{z \rho}} + O_2 \vec{\mathcal{F}_{z \rho}} = 0
\end{equation}

With $\vec{\mathcal{F}_{z \rho}} \equiv (\mathcal{F}_{z \rho}^{11} \dots \mathcal{F}_{z \rho}^{N_\rho N_z})$ and the matrices $O_1$ and $O_2$ are the grid-discretized versions of the differential operators
\begin{equation*}
\label{equ:matrix_operators}
O_1 \equiv \left[- 2  \p_z\right], \qquad  O_2 \equiv \left[ \p_\rho^2 + \frac{1}{\rho} \p_\rho - \frac{1}{\rho^2} + f(z) \p_z^2 +  f'(z) \p_z\right] 
\end{equation*}
In the end of the day one applies Runge-Kutta time evolution to system of equations \eqref{equ:evolution} in order to obtain the time dependent profile of the field at each point of the spatial grid. 

\begin{figure}[ht]
\center
\includegraphics[width=0.5\linewidth]{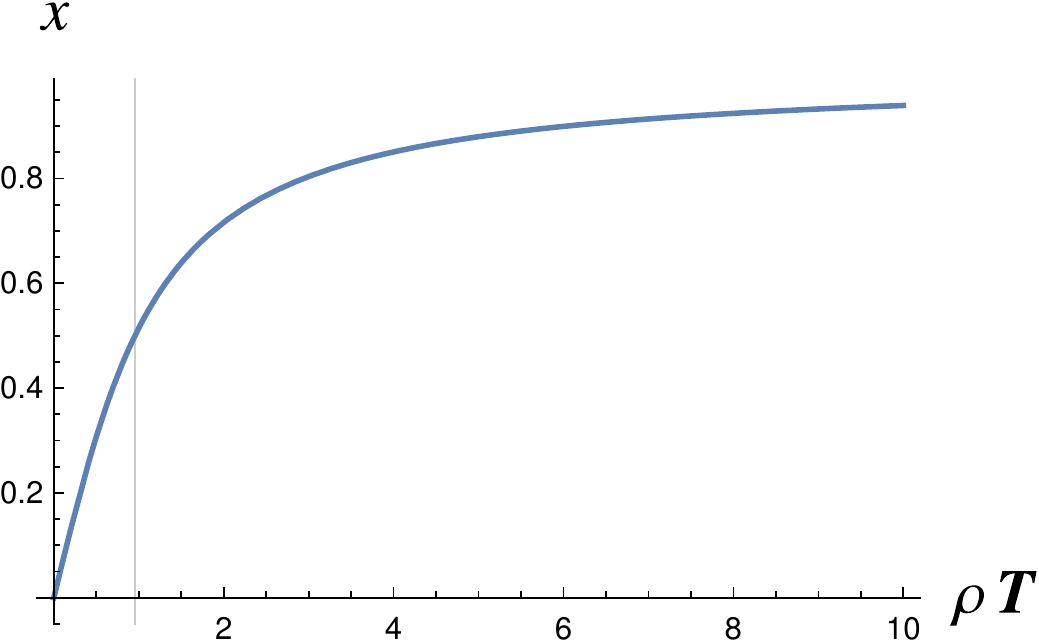}
\caption{\label{fig:REplot} 
The relation between the radial coordinate $\rho$ and the finite one $x$, which we use in the numerical simulation \eqref{equ:rx}, $c=0.95 \bm{T}^{-1}$}
\end{figure}

Before we turn to the numerical solution of the equations of motion \eqref{equ:EOMs}, we introduce the compact spatial coordinate (see Fig.\,\ref{fig:REplot})
\begin{equation}
\label{equ:rx}
x \equiv \frac{2}{\pi} \mathrm{arctan}\big( \frac{\rho}{c} \big),
\end{equation}
with the arbitrary parameter $c$, for which we use the value $c \approx 0.95 \bm{T}^{-1}$, which allows us to resolve the local quench profile \eqref{equ:dmu} in what follows. This coordinate transformation brings the infinite integration region in the radial coordinate $\rho \in [0,\infty)$ into the unit interval $x \in [0,1)$. We also rescale the holographic coordinate $z$ with horizon radius. Therefore, in the new coordinates the full integration region for the numerical problem is a unit square: $[0,1)_x \times (0,1]_z$.

In the problem under consideration we use the pseudospectral collocation method to represent the partial derivatives on the 2-dimensional grid of the size $31_x \times 24_z$. It is worth mentioning here that unlike similar calculations which were done in periodic lattices, we set 4 boundary conditions at all 4 boundaries of the integration region. We also introduce the Chebyshev descrete latices in both directions. We use the differentiation matrices in order to create the matrix operators \eqref{equ:matrix_operators}
discussed in i.e. \cite{boyd2001chebyshev,trefethen2000spectral} 
and implement the boundary conditions in the linear differential operators themselves, as discussed in \cite{Krikun:2018ufr}. We integrate the resulting linear system of time dependent ODEs using the standard evolution solver \texttt{NDSolve} in \textit{Wolfram Mathematica} \cite{Mathematica10}, the built in method \texttt{"EquationSimplification" -> "Residual"} proves to be useful.

\section{\label{app:QNMs} Quasinormal modes}
In order to study the spectrum of quasinormal modes, we need to find the values of $\omega$, for which the nontrivial solutions to the master equation \eqref{equ:ODE_QNMs} exist. Since it is a linear ordinary differential equation, the shooting method used in \cite{Bagrov:2016cnr,Krikun:2013iha,Andrade:2015iyf} suits the problem very well. 

We first factor out the oscillatory infalling wave \eqref{equ:BCH_QNM}
\begin{equation}
 \tilde{e}_\rho (z) \equiv  \left(1-\tfrac{z}{z_h} \right)^{i \omega z_h/3} e_\rho(z).
\end{equation}
Then expand the solution near the boundary (at $z=\delta z_b$) and the horizon (at $z=z_h - \delta z_h$) up to the 10-th order in $z_h \sim 10^{-2}$ and $z_b\sim 10^{-2}$. The expansion on the boundary depends on the two free parameters ($a_0, a_1$), of which the leading one, the source, we set to zero. At the horizon there is only one free parameter $b$, since the linearly independent outgoing branch of the solution has been excluded by the choice of the near horizon scaling \eqref{equ:BCH_QNM}. Using the obtained expansions we set the boundary conditions for the shooting procedure at the boundary 
\begin{align*}
{\tilde{e}_\rho}^\mathrm{boundary} (\delta z_b) &=  a_1 \, \delta z_b - \frac{1}{3}i \omega a_1 \, \delta z_b^2 + \dots, \\
\p_z \tilde{e}_\rho (\delta z_b) &=  a_1  - \frac{2}{3}i \omega a_1 \, \delta z_b + \dots.
\end{align*}
and at the horizon 
\begin{align*}
{\tilde{e}_\rho}^\mathrm{horizon} (z_h - \delta z_h) &=  b + b \frac{(3-2 i \omega) \omega^2 + 2 i \alpha^2 (3 i + \omega)}{3 \omega (3 i + 2 \omega)} \delta z_h + \dots, \\
\p_z {\tilde{e}_\rho}^\mathrm{horizon} (z_h - \delta z_h) &= - b \frac{(3-2 i \omega) \omega^2 + 2 i \alpha^2 (3 i + \omega)}{3 \omega (3 i + 2 \omega)} + \dots.
\end{align*}
Since the boundary value problem is overdetermined, there will be only descrete set of values $\omega_i(\alpha)$, for which the two solutions, originating from the boundary and the horizon could be smoothly connected  at arbitrary point inside the interval $z_0 \in (\delta z_b, z_h - \delta z_h)$:
\begin{equation}
\label{equ:match}
\frac{\p_z {\tilde{e}_\rho}^\mathrm{horizon} (z_0)}{ {\tilde{e}_\rho}^\mathrm{horizon} (z_0)} = \frac{\p_z {\tilde{e}_\rho}^\mathrm{boundary} (z_0)}{ {\tilde{e}_\rho}^\mathrm{boundary} (z_0)}, \qquad \forall z_0.
\end{equation}
In practice, for every value of $\alpha$ we check the matching relation \eqref{equ:match} in the range of complex values of $\omega$ and choose those were the relation holds for several values of $z_0$ (there are occasional cases when matching happens at particular value of $z_0$ only, and these must be excluded).

 \section{\label{app:EF} Eddington-Finkelstein coordinates}
% We use the Eddinton-Finkelstein coordinates 
% \begin{equation}
% \label{equ:v_coord}
% v = t - z^\ast, \qquad dz^\ast = dz f(z)^{-1}.
% \end{equation}

% The constant $v$ means $z^\ast$ grows to $\infty$ as $t$ increases. Given that near horizon $f(z) \sim (1-z)$ we get
% \begin{align}
% d z^\ast \sim \frac{dz}{1-z}
% z^\ast \sim - log(1-z)
% \end{align}
% Therefore $z^\ast \rar \infty$ means $log(1-z) \rar - \infty$ means $1-z \rar 0$, or simply $z \rar 1$. This proves that $v = t - z^\ast$ describes an infalling wave.

In order to make a transition from initial coordinates ($t,z$) to Eddington-Finkelstein ($v,z'$) we use the following rules.
\begin{align}
v(t,z) &= t - \int dz f(z)^{-1}, \\ 
\notag
z'(t,z) &= z
\end{align}
We can now rewrite the equations and functions $A_t, A_\rho$ in terms of the new variables. Note that we don't perform a tensor transformation on the vector $A_\mu$ 
\begin{align}
\p_t A(v,\rho,z') &= \frac{\p v}{\p t} \p_v A(v,\rho,z')+ \frac{\p z'}{\p t} \p_z' A(v,\rho,z') = \p_v A(v,\rho,z') \\
\notag
\p_z A(v,\rho,z') &= \frac{\p v}{\p z} \p_v A(v,\rho,z') + \frac{\p z'}{\p z} \p_{z'} A(v,\rho,z') = - f(z)^{-1}  \p_v A(v,\rho,z') +  \p_{z'} A(v,\rho,z')
\end{align}

Similarly
\begin{align}
\p_z^2 A(v,\rho,z') &= \p_{z'}^2 A(v,\rho,z') - 2 f(z)^{-1} \p_{z'} \p_v A(v,\rho,z') \\
\notag
& + f(z)^{-2} \p_v^2 A(v,\rho,z') - \p_{z}(f(z)^{-1}) \p_v A(v,\rho,z'). 
\end{align}

\bibliographystyle{JHEP-2}
\bibliography{local_quench}

\end{document}